\documentclass[conference]{IEEEtran}

\usepackage{fancyhdr}
\usepackage{graphicx}
\usepackage{multirow}
\PassOptionsToPackage{hyphens}{url}\usepackage{hyperref}
\usepackage{url}
\usepackage{subcaption}
\usepackage{calc}
\usepackage{amsthm}
\usepackage{amsmath}
\usepackage{algorithm}
\usepackage[noend]{algpseudocode}
\usepackage{algorithmicx}
\usepackage{xcolor}
\usepackage{multirow}

\begin{document}

\title{Towards Supporting Intelligence in 5G/6G Core Networks: NWDAF Implementation and Initial Analysis}

%
%
%

\author{Ali~Chouman, Dimitrios~Michael~Manias, and Abdallah~Shami\\
ECE Department, Western University, London ON, Canada\\
\{achouman, dmanias3, Abdallah.Shami\}@uwo.ca}


\maketitle

\begin{abstract}
Wireless networks, in the fifth-generation and beyond, must support diverse network applications which will support the numerous and demanding connections of today's and tomorrow's devices. Requirements such as high data rates, low latencies, and reliability are crucial considerations and artificial intelligence is incorporated to achieve these requirements for a large number of connected devices. Specifically, intelligent methods and frameworks for advanced analysis are employed by the 5G Core Network Data Analytics Function (NWDAF) to detect patterns and ascribe detailed action information to accommodate end users and improve network performance. To this end, the work presented in this paper incorporates a functional NWDAF into a 5G network developed using open source software. Furthermore, an analysis of the network data collected by the NWDAF and the valuable insights which can be drawn from it have been presented with detailed Network Function interactions. An example application of such insights used for intelligent network management is outlined. Finally, the expected limitations of 5G networks are discussed as motivation for the development of 6G networks.
\end{abstract}


\section{Introduction}
%
%
%
%
\IEEEPARstart{T}{he} telecommunications industry has sparked a dramatic transition to novel and improved high-speed wireless communications architectures in industry and society. The design and operation of 5G and Beyond (5G+) networks is a tightly woven cooperation of developments in both the 5G Core and 5G radio networks that has led the charge for fast-paced development in the communications industry. The 5G+ concept has become a critical tool in the introduction and development of Industry 4.0, a paradigm shift of modern wireless communications systems to true, digital economies \cite{rommer20195g}.

The 5G architecture is comprised of the 5G Core (5GC) network, the new Radio Access Network (RAN), and its newly supported New Radio (NR). The Third Generation Partnership Project (3GPP) outlined the design of the 5G Core to implicitly and explicitly support new architectural features, such as a service-based architecture (SBA), consistent user experience, improved Quality-of-Service (QoS), enhanced machine-to-machine communication services, adaption to cloud-native technologies, and edge computing access. 5G defines three service grades, where each strata defines its own special requirements to adhere to customers' business models: Ultra-Reliable Low Latency Communications (URLLC), Massive Machine-Type Communications (mMTC), and Enhanced Mobile Broadband (eMBB) \cite{ETSI_NWDAF}. 

The use of AI in 5G+ networks is one of the defining characteristics of this paradigm-shifting technology. According to reported statistics, by 2025, it is projected that the telecom industry will invest USD 36.7B in AI through software and hardware investments as well as AI services \cite{IntelAI}. The operational benefits of AI in 5G+ networks consider the added value the AI system provides in terms of the management and orchestration of networks \cite{ManiasNetowrk20}\cite{ManiasGlobecom}. One of the main benefits of AI is the ability to take proactive and predictive measures to ensure the optimization of network performance. Some methods of network performance optimization include the reduction of power consumption through enhanced algorithmic performance, the maximization of throughput through optimal traffic routing and infrastructure placement, as well as the ability to support an increasingly dense number of users \cite{deepsig}\cite{HawiloNetwork}.\par

The envisioned AI-enabled network will consist of intelligent agents being fed data related to the network, including network measurements and statistics, resource utilization, and traffic patterns and conducting inferencing to provide network automation through Management and Orchestration (MANO) tasks such as resource optimization and system reconfiguration \cite{IntelAI}\cite{HawiloJSAC}. However, one of the main challenges plaguing AI implementations across all fields relates to the availability of high-quality data. In order to effectively collect the required data to build and train AI agents, data collection interfaces need to be deployed throughout the network and constantly be monitored \cite{5GPPP}. To this end, the Network Data Analytics Function (NWDAF) has been proposed by 3GPP as a solution to this problem to be directly implemented in the 5G+ core network as a key network function.\par

The work described in this paper addresses the practical development of the NWDAF and considers its integration into an operational 5G core implemented using open-source software, including Open5GS \cite{Open5GS} and UERANSIM \cite{aligungr}. The main contributions of this paper can be summarized as follows:
\begin{itemize}
    \item Initial analysis into the type of core network function data that can be collected by the NWDAF.
    \item Presentation of insights which can be drawn using the collected core network data.
    \item Discussion as to how the NWDAF can be used to influence MANO activities such as core network function placement.
    \item An outlook into the state of future networks, the expected limitations of 5G, and the motivation sparking the initial discussion of 6G networks.
\end{itemize} \par 

The remainder of this paper is organized as follows. Section II considers background information related to the 5G Core and the NWDAF. Section III presents a use case highlighting key insights obtained from the analysis of NWDAF-collected 5G core network data and its application to the MANO of 5G+ networks. Section IV discusses the vision and requirements of 6G networks, as well as the expected limitations of 5G networks motivating their initial conceptualization and development. Finally, Section V concludes the paper and discusses opportunities for future work.

\section{Background}

\subsection{5G Core}
The 5G Core is composed of various Network Functions (NFs) with their individually associated microservices and responsibilities. The 3GPP intended for the 5G Core to bring about a mindset shift from evolving architectures into standalone, access-independent structures. For example, the 5G Core, by design principle, does not provide backwards compatibility for any previous generations of RANs (\emph{e.g.,} GSM, LTE). Instead, the 5GC consists of a new set of interfaces that are intended for core network-radio network interactions. In terms of the 5G RAN specifications, the 3GPP defined two architectural variants which combine the LTE and the 5G NR: the non stand-alone architecture (NSA) and the stand-alone architecture (SA). The key difference is that the NSA aims to maximize the reusability of 4G architectures by relying on LTE radio access for signaling between UE devices and the network. Specifically, it consolidates an enhanced EPC network to support 5G in the more recent deployments \cite{rommer20195g}.\par

At the core of 5GC, NFs provide the functionality for establishing sessions and forwarding data to and from mobile User Equipment (UE) devices. Some key NFs and their operations are detailed to provide a brief summary of the 5G Core functionalities. The Access and Mobility Management Function (AMF) interacts with the UE devices and the RAN, and is involved in most 5G signalling calls. As well, the AMF supports activation for devices in idle mode. The Session Management Function (SMF) manages end user device sessions, including their establishment, modification, release, and IP address allocation. The SMF also interacts with other NFs to select and control different User Plane Function (UPF) instances over the network. This control allows it to configure traffic steering and enforcement in UPFs for individual sessions. The UPF processes and forwards user data and is controlled by the SMF. In addition, the UPF connects to external IP networks to act as anchor points, hiding mobility. The Unified Data Management Function (UDM) accesses user subscription data stored in the Unified Data Repository (UDR), a database containing network/user policies and associated data. Finally, the Authentication Server Function provides authentication services for a specific device, utilizing credentials from the UDM \cite{brown2017service}. \par

As an underlying function solely responsible for data analytics and network learning, the NWDAF represents operator-managed network analytics as a logical function \cite{ETSI_NWDAF}. The NWDAF provides slice-specific network data analytics to any given NF. As well, the NWDAF provides network analytics information to NFs on a network slice instance level. The function also notifies NFs with slice-specific network status analytic information for any that are subscribed to it. NFs may also collect network status analytic information directly from the NWDAF. In the 5G Core, both the Policy Control Function (PCF) and the Network Slice Selection Function (NSSF) are consumers of network analytics. The PCF may use that data in its policy decisions, and the NSSF may use the load-level information provided by the NWDAF for slice selection. \par

\subsection{NWDAF}
The NWDAF architecture is designed to aid policy and decision-making for NFs in the control plane and supports some important services for a given NF service consumer. Industrial NWDAF solutions typically have an N23 interface and an N34 interface as reference points to the PCF and the NSSF, respectively. As well, 5G edge computing use cases allow the NWDAF to aid the SMF in routing decisions. As the central point of network analytics, the NWDAF enables operators to capture non-SBI data in addition to SBI data as standalone 5G deployments become more prevalent  \cite{radcom_2021}.

As of December 2021, the NWDAF provides five different NF services: \emph{AnalyticsSubscription}, \emph{AnalyticsInfo}, \emph{DataManagement}, \emph{MLModelProvision}, and \emph{MLModelInfo}. The \emph{AnalyticsSubscription} service notifies the NF consumer instance of all analytics subscribed to the specific NWDAF service. The \emph{AnalyticsInfo} service enables the NF consumer to request and retrieve network data analytics from the NWDAF. As well, it enables the NWDAF to request analytics context transfers from another NWDAF if necessary. The \emph{DataManagement} service allows an NF consumer to subscribe to receive data or historical analytics (interpreted as data); if the data is already defined in the NWDAF, the subscription is updated. The \emph{MLModelProvision} service enables an NF consumer to receive notifications when an ML model, matching subscription parameters, becomes available. Finally, the \emph{MLModelInfo} service enables an NF consumer to request and retrieve ML model information from the NWDAF \cite{3gppR16}.

Industrial NWDAF implementations provide closed-loop automation for third-party NFs and solutions inside the 5G Core. In particular, these NWDAFs are intended for continuous monitoring of every NF, network slice, and UE device and use a variety of KPIs to measure network performance. The real-time KPIs can be used to automate network issue resolution, while predictive analytics can be used to predict those network issues in the future. Predictive analytics may also provide anomaly detection to be used for automating mitigation \cite{radcom_2021}. \par

\section{Case Study: NWDAF Implementation and Analysis}
The following case study will explore the various insights and conclusions drawn from network-generated data from a 5G Core Network. The analysis conducted in this case study is an example of how the NWDAF can leverage data to provide meaningful insights to enhance the MANO of core network functions. Specifically, this case study will analyze control packets generated during the instantiation of the network core. Through these control packets, various statistics such as the size and number of packets per protocol will be displayed. Additionally, an in-depth exploration of the Binding Support Function (BSF) and its interaction with the Network Repository Function (NRF) will be discussed. Using both analyses, a recommendation can be made regarding the placement of the BSF in relation to the NRF. \par

\begin{figure*}[!htbp]
\centerline{\includegraphics[width=1.60\columnwidth]{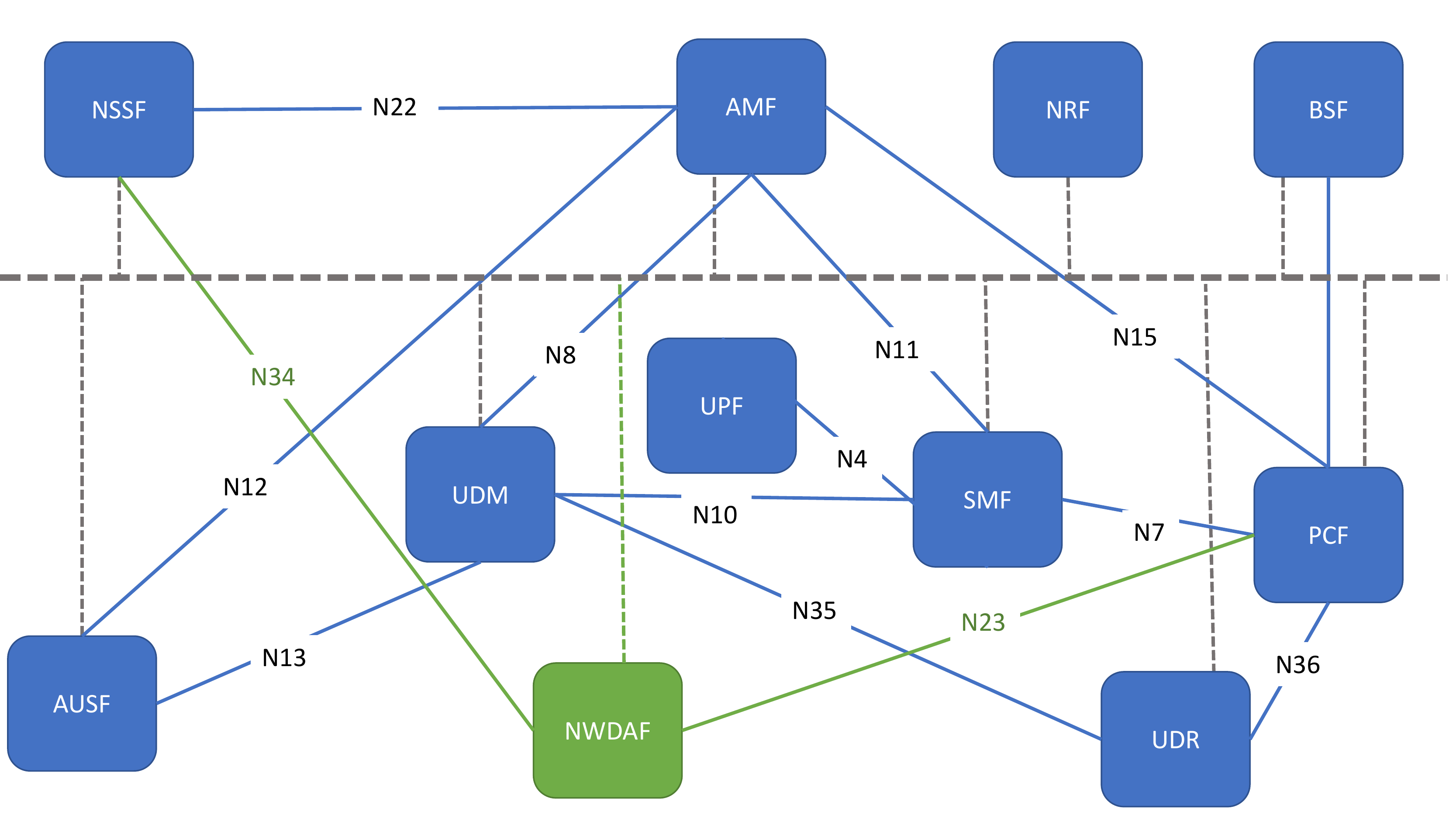}}
\caption{5G Core Service Based Architecture}
\label{5G_SBA}
\end{figure*}

The collected data for this case study was generated through Open5GS, an open-source project providing network functionalities for building private 5G networks \cite{Open5GS}. The 5G standalone implementation was used for the system model leveraging both the Service-Based Architecture (SBA) and following the Control and User Plane Separation (CUPS) scheme, as described by 5G network standardization efforts led by the 3GPP \cite{3gppR16}. UERANSIM, an open-source state-of-the-art 5G UE and RAN implementation, was used to complete full operation of the 5G Core with connected devices \cite{aligungr}. Figure \ref{5G_SBA} outlines the various core network functions which were operational during the data collection phase. Additionally, this figure illustrates how the proposed NWDAF fits in the 5G Core with its associated interfaces (depicted in green). In this figure, the reference point architecture, presented by solid lines, illustrates the point-to-point interaction between core network functions, whereas the SBA is illustrated by the dashed lines. Through the SBA, the NWDAF is able to collect data and statistics about all other authorized core network functions without having an explicit point-to-point reference defined. The data collection phase ran for 138 minutes and, as previously mentioned, exclusively considered the control signalling between the various network functions in the control plane, not including any GPRS Tunneling Protocol (GTP) traffic from the UE and the RAN. The data for this case study is publically available \cite{data}.  \par

\begin{figure}[!htbp]
\centerline{\includegraphics[width=0.95\columnwidth]{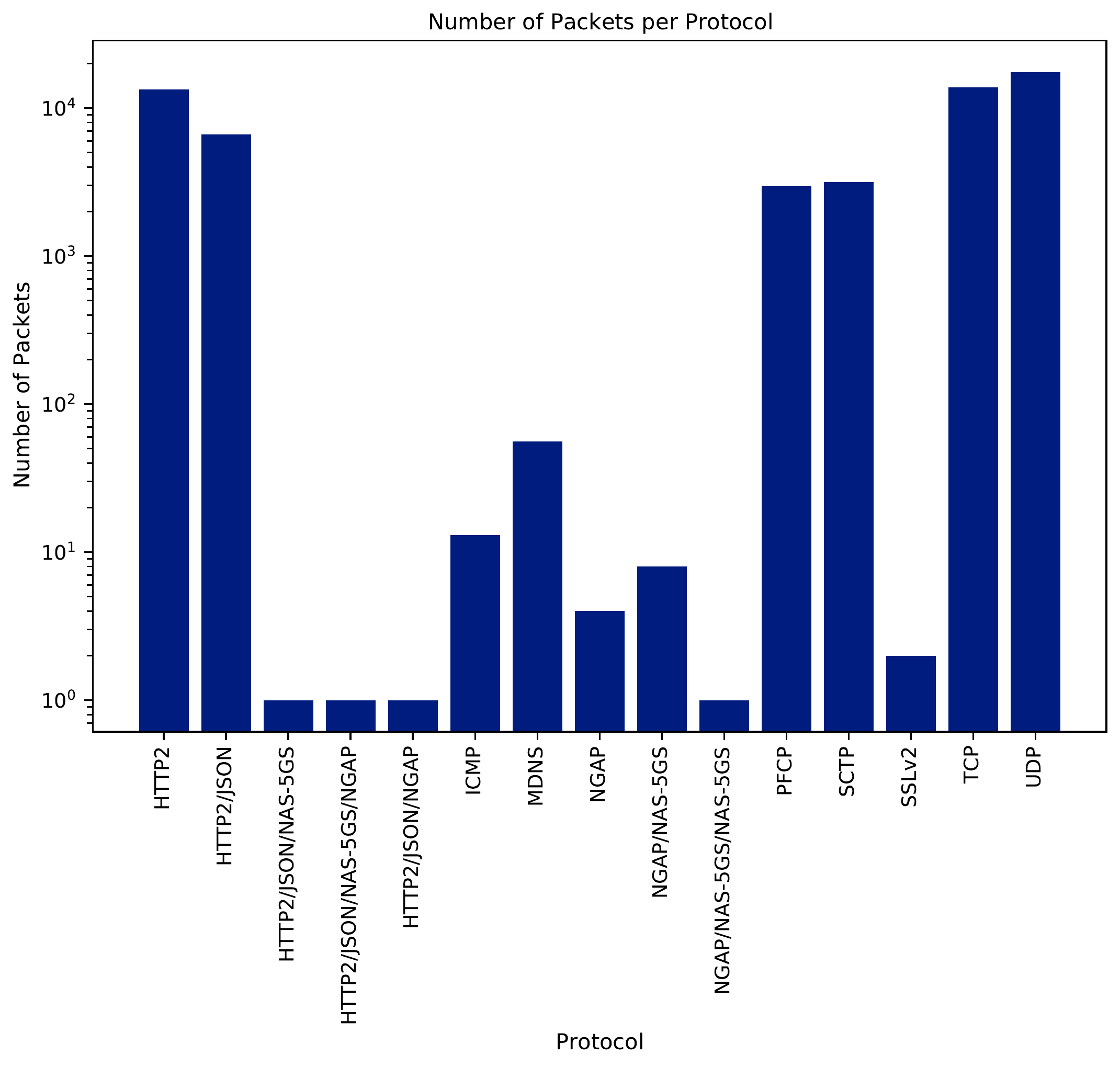}}
\caption{Number of Packets per Protocol}
\label{packets_protocol}
\end{figure}

The first result presented in Fig. \ref{packets_protocol} considers the total number of packets associated with each observed protocol throughout the duration of the data collection phase. As seen through this figure, the overwhelming majority of packets utilize the TCP protocol, something which is expected considering the NFs communicate with each other through REST APIs leveraging the HTTP/2 protocol. RESTful SBA procedures can be categorized into Service Registration, Service Discovery, and Session Establishment. It should be noted that the three NGAP protocols have been introduced in 5G and are used in communications between the gNB and the Access and Mobility Function.  \par

In addition to the NGAP protocols, which are prevalent in UE registration and de-registration, the Packet Forwarding Control Protocol (PFCP) is paramount to formalizing the interactions between 5G Core NFs, specifically between the SMF and the UPF through the N4 interface. Albeit infrequent in the generated network traffic when compared to other protocols, PFCP is used in signalling procedures in the Control Plane for network attachment and in the User Plane for IPv4/IPv6 packet forwarding with the wireless RAN and the PDU \cite{mohamed2021performance}.

\begin{figure}[!htbp]
\centerline{\includegraphics[width=0.95\columnwidth]{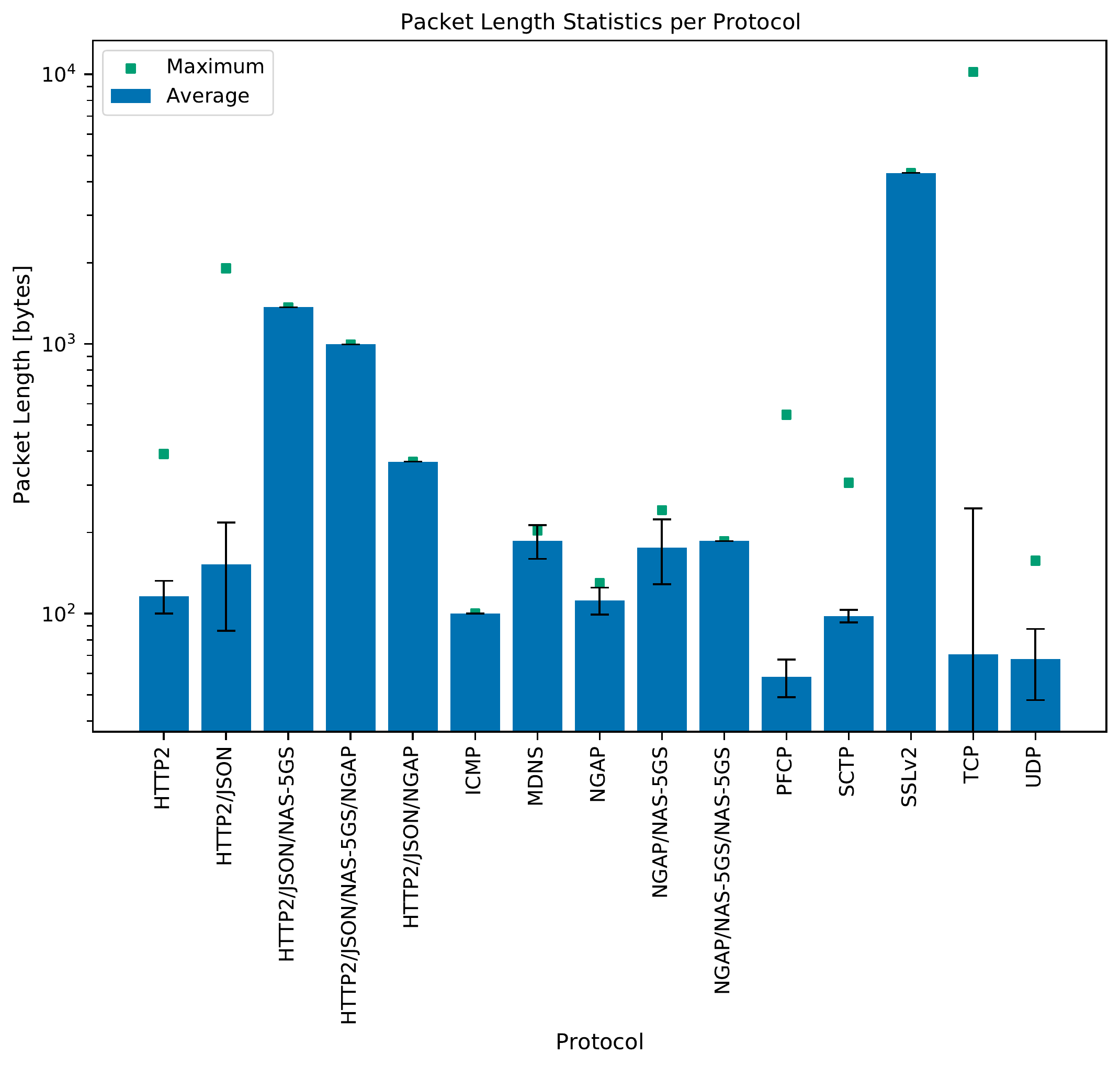}}
\caption{Packet Length Statistics per Protocol}
\label{packet_stats}
\end{figure}

The next stage in this analysis considers the average size of each protocol’s packets along with statistics such as the standard deviation and maximum packet size as presented in Fig. \ref{packet_stats}. Through this figure, it can be seen that the largest packet sizes are attributed to the SSL protocols. However, given the volume of SSL packets presented in the previous results, these packets are infrequent. Considering both presented results, it is evident that the focus of this analysis should be on TCP packet signalling as they have the greatest volume and significant size compared to the other protocols.  \par

\begin{figure}[!htbp]
\centerline{\includegraphics[width=0.95\columnwidth]{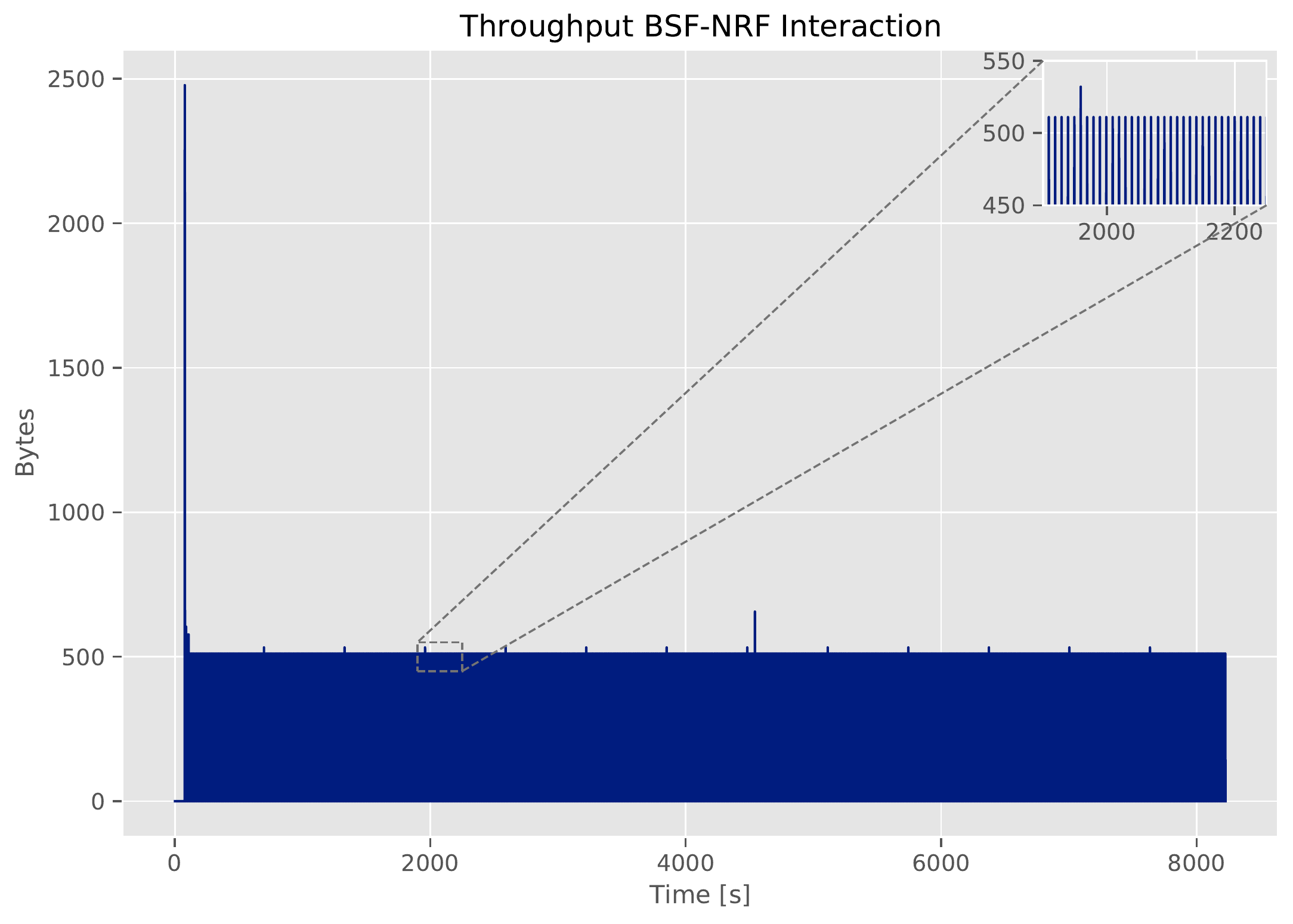}}
\caption{Bytes per Second Exchanged between BSF and NRF}
\label{TCP_20_tp}
\end{figure}

The following results pertain to the interaction between the BSF and the NRF. This interaction was selected to further explore the trends in TCP control packets and provide a meaningful recommendation based on the volume and frequency of data exchanged between these NFs. Figure \ref{TCP_20_tp} considers all TCP control packets exchanged between these two functions and compares the size of the packet to the time at which it was sent, effectively providing bi-direction link throughput for this interface. This figure shows a clear spike in packet size near the beginning, followed by a constant packet size for the remainder of the data collection stage. The zoomed-in portion of the graph shows that the packets are transmitted periodically with minor variations due to signalling processes. \par

\begin{figure}[!htbp]
\centerline{\includegraphics[width=0.95\columnwidth]{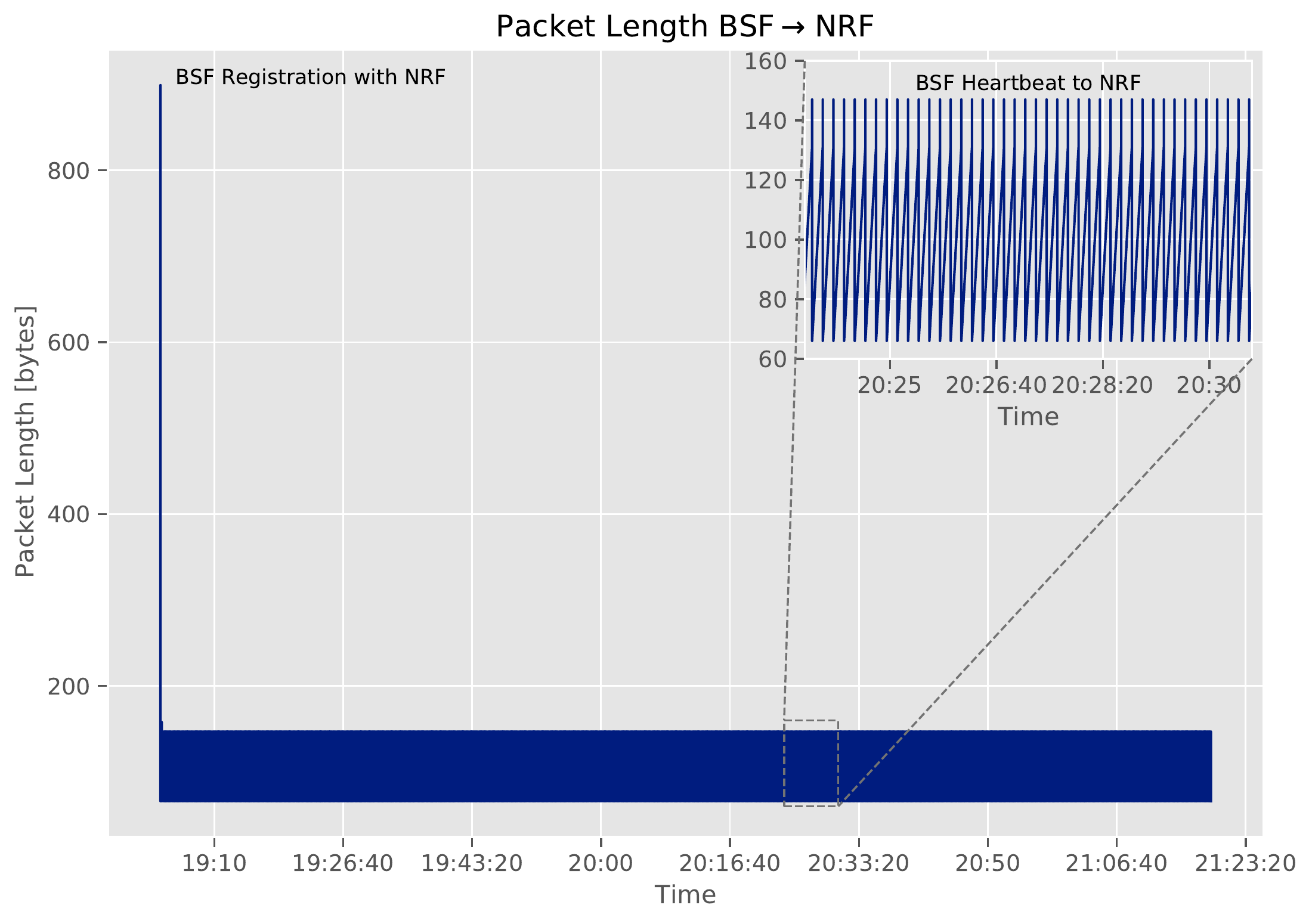}}
\caption{Length of Packets Sent from BSF to NRF}
\label{BSF_to_NRF}
\end{figure}

To further explore these results and translate them to observable 5G NF events, a more intuitive analysis is done regarding the one-way communication of the BSF with the NRF as seen in Fig. \ref{BSF_to_NRF}. In this figure, the TCP packets with the BSF as the source and the NRF as the destination are displayed, and text annotations have been used to highlight key NF events. As seen in this figure, the initial spike in packet size is attributed to the BSF registering with the NRF. In HTTP/2-based communications, this interaction corresponds to a request/response fetch in the SBI, as opposed to a subscribe/notify callback (\emph{e.g.,} the SMF subscribing to the NRF for notifications when other NFs go down). During this registration process, the BSF is required to send all its functional information to the NRF, resulting in the increased packet size. The second major NF event is the BSF heartbeat which occurs every 10 seconds and makes the NRF aware of any changes in its status (\textit{i.e.,} registration, load). As illustrated in the zoomed-in portion of the graph, two packet size values emerge; the greater packet size is associated with the PATCH request used to perform the heartbeat, whereas the lesser packet size is associated with the acknowledgement of received information from the NRF in response. It is important to note that the PATCH request, partially updates the network resource, compared to a PUT request, which completely replaces the resource addressed by the URI with the JSON-formatted payload of the request. \par

\begin{figure}[!htbp]
\centerline{\includegraphics[width=0.95\columnwidth]{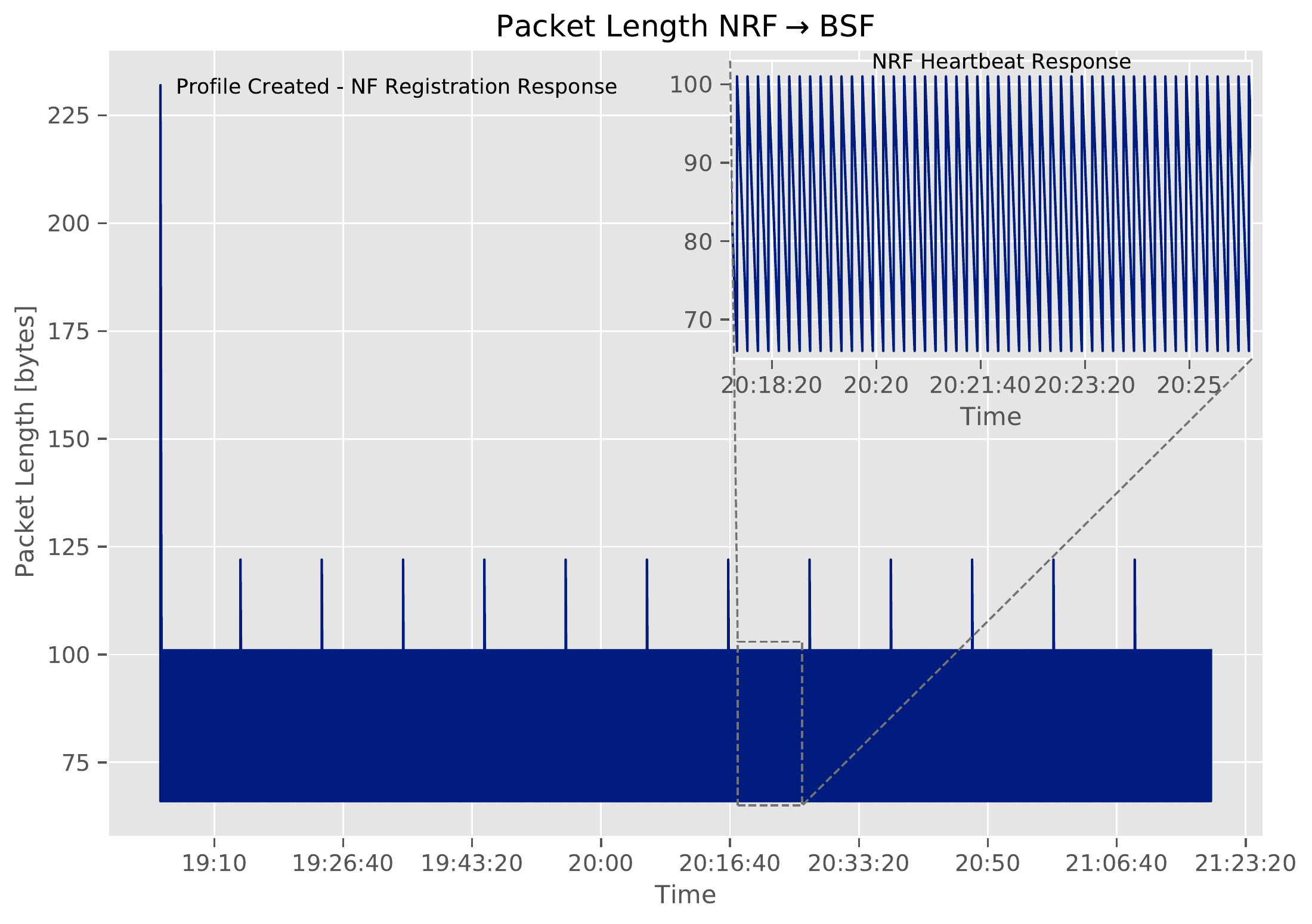}}
\caption{Length of Packets Sent from BSF to NRF}
\label{NRF_to_BSF}
\end{figure}

Equivalently, Fig. \ref{NRF_to_BSF} presents the one-way communication of the NRF with the BRF. As labelled through the annotation, the first major spike corresponds to the response sent when the NF has been registered (\emph{NF\_REGISTERED}) and the profile has been created. As expected, when compared with the initial request seen in Fig. \ref{BSF_to_NRF}, the response is significantly smaller. Furthermore, when looking at the zoomed-in portion of the graph outlining the response to the BSF heartbeat, there are once again two distinct packet sizes that emerge. The smaller of the two sizes corresponds to a simple acknowledgement, whereas the larger size corresponds to the response of the heartbeat. As outlined in the NRF schema, if no significant change has been made to the status of the function, the response to the heartbeat is a packet with an empty body; however, if there were to be a significant change to the function status, such as the signal value \emph{NF\_DEREGISTERED}, this response’s body would contain the latest updated information.

Given the results presented in this case study, the NWDAF could be tasked with recommending a placement decision for another instance of the BSF function. An industrial NWDAF solution utilizes similar policy decision-making in the context of the PCF and the NSSF \cite{radcom_2021}. The PCF should take input from the NWDAF to allocate resources and steer traffic policies for dynamic network slices, and the NSSF should gather load-level information from the NWDAF for the purpose of slice selection. As illustrated, the initial registration process with the NRF results in the packet size spike, whereas the remainder of its interaction with the NRF is a set of periodic heartbeats of much smaller packet size. For this reason, the co-location of these network functions is likely not required as the amount of control information exchanged between them is limited. Future work with the BSF will explore its interaction with the PCF as it is responsible for communicating with the PCF to partially update binding information for PDU sessions which are set to binding level endpoint \emph{NF\_SET}. \par

\section{Future Outlook}

Efficient operation is at the forefront of future 5G deployment and maintenance considerations, and mobility is a key metric/KPI that the NWDAF must follow and learn about within the network. Control plane NFs can predict this mobility, provided that the analytics operations are efficient and intelligent (\emph{i.e.,} leads to making proactive decisions). Data required for mobility prediction can be streamed by the AMF to the NWDAF via the data collection API and can identify aperiodicity in mobility patterns. On the user plane side, a UPF area prediction service can be implemented based on UE location, capacity/availability, and distance \cite{jeong2021mobility}.

While 5G is currently being developed and NSPs are beginning initial deployment, and integration, discussion of 6G networks is already on the horizon. The main motivation behind this discussion is the rapidly increasing number of connected devices, a trend that is expected to continue for the foreseeable future as I/IoT frameworks continue to develop and expand. To this end, there is a concern that 5G networks will not be able to meet the demands of future use cases, which are projected to require transfer rates in the order of Tb/s, latency on the order of microseconds, as well as increased connection density due to a multitude of deployed sensors. Additionally, future use cases ranging from Augmented/Virtual Reality (A/VR) and Robotics to Holograms and Teleoperation are expected to disrupt existing networks and test their extraneous, boundary use cases \cite{giordani2020toward}. \par

To address the expected limitations and shortcomings of 5G networks, research has begun into methods of expanding the capacity and capability of future networks to ensure that future demands can be met. Some proposed lines of research include the exploration of new frequency bands (THz) as well as distributed and federated intelligence throughout the network. As user behaviour and habits change and evolve, networking practices must also follow suit. 6G presents a revolutionary opportunity to scale up the presence of intelligence in networks and ultimately enable a plethora of future use cases and applications; however, this is not a trivial task.  \par
As 6G networks take shape, AI will be deeply integrated into the network, more than just through a core network function. As intelligence gets distributed through the system, so do the privacy and security risks associated with it. These risks can range from data poisoning at edge nodes to system-wide model drift, each with its own intricacies and nuances which must be addressed. Additionally, with more data distributed at the edge, data privacy is paramount to ensuring public safety considering critical services such as emergency, finance, and transportation will be in jeopardy of being compromised. As such, it will become increasingly important to consider model maintenance as an integral part of the ML/AI life cycle to ensure future networks' safe and secure performance.
\par

6G networks must fully realize the revolutionizing Industry 4.0 that started with 5G networks. In particular, it is the digital transformation of physical manufacturing systems and IoT services. IoT-based diagnostics will enhance maintenance and operation of machine communications, prioritizing cost-effectiveness and flexibility. In Industry 4.0, automation requires reliable and synchronous communication systems that 6G is situated to address through the aforementioned disruptive technologies \cite{zhang20196g}.\par
The various use cases of the NWDAF will aid in addressing the AI-related challenges in 6G networks. With the advent of low-power requirements for IoT devices, AI model training could consider new specifications recommended by NWDAF data, based on federated learning (\emph{e.g.,} learning at edge devices). As well, AI use cases guided by the NWDAF data analytics must address the lack of bounding performance in 6G networks. In contrast to the previous challenge, system design must consider worst-case scenario network events while providing a minimum acceptable QoS/performance guarantee; however, due to non-linear characteristics of such related problems, it may be infeasible for AI approaches regardless of their effectiveness in real-time inferences \cite{shafin2020artificial}. 
 
\section{Conclusion}
The adoption and integration of intelligence in 5G networks has the potential to revolutionize our current networking practices. Perhaps the greatest potential lies in the Network Data and Analytics Function (NWDAF) proposed by 3GPP. This function will collect a plethora of information and statistics on the operation of the network ranging from high-level data such as slice level information to very specific data related to a single NF. This will ultimately enable a holistic view of the network and will enable enhanced Management and Orchestration (MANO) capabilities. Throughout this paper, we have presented a case study that outlined an analysis of NWDAF-collected core network function data from an Open5GS and UERANSIM implementation. An initial analysis into this data and the potential insights that can be drawn from it were illustrated. In this case study, 5G core function control messages were considered; specifically, the interaction between the Network Repository Function (NRF) and the Binding Support Function (BSF) was examined. \par

Future work in this area will consider the impact of the NWDAF on 5G networks and continue to explore data generated from 5G core network functions. As mentioned in the results section, a study on the interaction between the BSF and the Policy and Control Function (PCF) will be a focus. Finally, the development of advanced analytics models will be considered using the generated data for use cases such as proactive network management and forecasting.

\bibliographystyle{unsrt}
\bibliography{sample}

\end{document}